\documentstyle[11pt,aaspp4]{article}

\slugcomment{Accepted for publication in the Astrophysical Journal}

\lefthead{}
\righthead{ISO-SWS spectroscopy of NGC 1068}

\begin{document}

\title{ISO-SWS spectroscopy of NGC 1068
\footnote{Based on observations with ISO, an
   ESA project with instruments funded by ESA Member States (especially the PI
   countries: France, Germany, the Netherlands and the United Kingdom) with the
   participation of ISAS and NASA.}
   }

\author{D.~Lutz\footnote{Max-Planck-Institut f\"ur extraterrestrische Physik,
   Postfach 1603, 85740 Garching, Germany},
   E.~Sturm$^2$,
   R.~Genzel$^2$,
   A.F.M.~Moorwood\footnote{European Southern Observatory,
     Karl-Schwarzschild-Stra\ss\/e 2, 85748 Garching, Germany},
   T.~Alexander\footnote{Institute for Advanced Study, Olden Lane,
         Princeton, NJ 08540, USA},
   H.~Netzer\footnote{School of Physics and Astronomy and Wise Observatory,
      Raymond and Beverly Sackler Faculty of Exact Sciences, Tel Aviv
      University, Ramat Aviv, Tel Aviv 69978, Israel},
   A.~Sternberg$^5$
   }

\begin{abstract}
We present ISO-SWS spectroscopy of NGC 1068 for the complete wavelength 
range 2.4
to 45$\mu$m at resolving power $\sim$1500. Selected subranges have been
observed at higher sensitivity and full resolving power $\sim$2000. We 
detect a total of 36 emission lines
and derive upper limits for 13 additional transitions. Most of the observed 
transitions are fine structure and recombination lines originating in the
narrow line region (NLR) and the inner part of the extended emission line region.

We compare the line profiles of optical lines and reddening-insensitive 
infrared lines to constrain the dynamical structure and extinction
properties of the narrow line region.
The most likely explanation of the considerable differences found is
a combination of two effects. (1) The spatial structure of the NGC\,1068 
narrow line region is a combination of a highly ionized outflow 
cone and lower excitation extended emission. (2) Parts of the narrow line
region, mainly in the receding part at velocities above systemic, are subject 
to extinction that is significantly suppressing optical emission from
these clouds. Line asymmetries and net blueshifts remain, however, even for
infrared fine structure lines suffering very little obscuration. This may
be either due to an intrinsic asymmetry of the NLR, as perhaps also suggested 
by the asymmetric radio continuum emission, or due to a very high column 
density obscuring component which is hiding part of the narrow line region
even from infrared view. 

We present detections and limits for 11 rotational and ro-vibrational emission 
lines of molecular hydrogen (H$_2$). They arise in a dense molecular medium at 
temperatures of a few hundred Kelvin that is most likely closely related to
the warm and dense components seen in the near-infrared H$_2$ rovibrational 
transitions, and in millimeter wave tracers (CO, HCN) of molecular gas. Any 
emission of the putative
pc-scale molecular torus is likely overwhelmed by this larger scale
emission.

In companion papers we use the SWS data to derive the spectral energy 
distribution
emitted by the active nucleus of NGC\,1068 (\cite{alexander00}), to put
limits on infrared emission from the obscured broad line region 
(\cite{lutz00}), and discuss
the continuum and its features in conjunction with SWS spectra of other galaxies
(\cite{sturm00}).
\end{abstract}


\section{Introduction}

NGC\,1068 is one of the nearest and probably the
most intensely studied Seyfert 2 galaxy. Observations in all wavelength
bands from radio to hard X-rays have formed a uniquely detailed picture
of this object.
NGC 1068 has played a key role in the development of unified scenarios
for Seyfert 1 and Seyfert 2 galaxies (\cite{antonucci85}),
in the study of molecular gas in the nuclear region of Seyferts
(e.g. \cite{myers87}; \cite{tacconi94}),
and in elucidating the importance of star formation activity coexistent
with the AGN, both on
larger (e.g. \cite{telesco88}) and smaller (\cite{macchetto94}; \cite{thatte97})
scales. NGC 1068 hosts a prominent Narrow Line Region (NLR) that is 
approximately cospatial with a linear radio source with two lobes
(\cite{wilson83}). The narrow emission line region has been extensively
characterized from subarcsecond clouds probed by HST (\cite{evans91};
\cite{macchetto94}), the  
$\approx 5$ arcseconds of the NLR hosting most of the line flux
(e.g. \cite{walker68}; \cite{shields75}; \cite{cecil90}), and
the ionization cone and extended emission line region 
(\cite{pogge88}; \cite{unger92}) extending to radii of at least 30\arcsec\/ 
(1\arcsec\/ = 72 pc at the distance of 14.4 Mpc, \cite{tully88}). 
The velocity field is complex, with an ensemble of rapidly
moving clouds dominating the inner arcseconds and a more quiescent rotation
pattern prevailing at larger radii (e.g. \cite{walker68}; \cite{alloin83};
\cite{meaburn86}; \cite{cecil90}). While most of the excitation of the narrow
line region and the extended emission line region is likely through 
photoionization
by the central AGN (\cite{marconi96}), high resolution observations suggest 
kinematic disturbance and possibly shock excitation of regions close to the 
radio outflow (e.g. \cite{axon98}).  

With ESA's {\it Infrared Space Observatory} ISO, sensitive mid-infrared 
spectroscopy of AGNs became possible, with detections of a broad range 
of low- and high-excitation fine structure lines, recombination lines,
and pure rotational lines from such sources
(\cite{moorwood96}; \cite{sturm99}; \cite{alexander99})).
Model predictions of the mid-IR spectra of AGN had been obtained prior to 
ISO (e.g. \cite{spinoglio92}), but observations were restricted by limited
sensitivity and focussed primarily on the continuum emission and broad
features rather than emission lines. 

In this paper, we present the ISO-SWS
spectra of NGC\,1068 and draw conclusions on the structure of
the narrow line region that can be obtained mainly from the comparison
of optical and reddening-insensitive infrared lines, and discuss the 
nature of the mid-infrared molecular hydrogen emission.
The ISO-SWS data of NGC\,1068 are analysed further in several 
companion papers.
Alexander et al. (2000) use photoionization modelling based on the ISO fine 
structure line set and other NLR lines to model the shape of the AGN's spectral
energy distribution.
Lutz et al. (2000) analyze limits on emission from the obscured broad line 
region.
Finally, Sturm  et al. (2000) discuss continuum energy distribution and 
features of NGC\,1068 in conjunction with ISO-SWS spectra of other galaxies.

Our paper is organised as follows. In  \S 2 we discuss the ISO-SWS observations
and data reduction. \S 3 presents results and implications of the
density of the narrow line region. \S 4 uses infrared line profiles in 
comparison to optical ones to constrain the structure of the narrow line region.
We discuss the mid-infrared molecular hydrogen emission in \S 5 and summarize
in \S 6. 

\section{Observations and data reduction}
\label{sect:obs}

We have 
used the Short Wavelength Spectrometer SWS (\cite{degraauw96}) on board
the {\it Infrared Space Observatory} ISO (\cite{kessler96})
to observe the nuclear region of NGC\,1068. In 
Table~\ref{tab:listobs} we present a log of our observations. We carried
out observations in the
SWS01 mode which provides a full 2.4 -- 45$\mu$m scan at slightly reduced 
spectral resolving power, as well as
 observations in the SWS02 and SWS06 modes targeted at full resolution 
observations of individual lines or short ranges.
Because of the large width of the emission lines in the NGC\,1068 narrow line
region, we have mostly relied on the SWS06 mode which can be set up to 
provide
wider continuum baselines than standard SWS02 line scans. We supplement
observations from our ISO guaranteed and open time with serendipitous 
information on some fine structure lines obtained in another ISO project
(PI G.~Stacey), the main results of which are to be presented elsewhere.

Table~\ref{tab:listobs} includes also the position angle of the long
axis of the SWS apertures for the various observations. Our 
pointing was always centered on the nucleus of NGC 1068, but the SWS apertures
range from 14\arcsec$\times$20\arcsec\ at short wavelengths to 
20\arcsec$\times$33\arcsec\ at the longest wavelengths (\cite{degraauw96}).
At the long wavelengths, we partly include the $\sim$15\arcsec\ 
radius ring of star forming regions encircling the nucleus of NGC 1068.   
The apertures were always oriented in approximately north-south (or south-north)
direction, with position angles between -11\arcdeg\/ and -23\arcdeg\/.

We have analyzed the data using the SWS Interactive Analysis (IA) system
(\cite{lahuis98}; \cite{wieprecht98})
and calibration files of July 1998. A preliminary account of part of
the observations is given by Lutz et al. (1997). Since then, calibration
files have been updated for wavelength calibration and in particular
with respect to the  SWS relative spectral 
response function, leading to more reliable intercalibration
between the `AOT bands' forming a full SWS spectrum. 
Our data reduction started in the standard way and 
continued with steps of (interactive) dark current 
subtraction and matching up- and downscans. We eliminated data from
those detectors of band 3 that were most noisy during a particular revolution,
and from interactively identified regions
with single detector signal jumps in bands 1 and 2, and simultaneous 
12-detector signal
jumps in band 3. After relative spectral response correction and flux 
calibration, we `flatfielded' the 12 detectors of a band to a consistent level,
corrected for the ISO velocity, and extracted the AAR data product. Redundant 
scans
of the same line were shifted to a consistent level. Single-valued spectra were
produced by kappa-sigma clipping the AAR dot cloud and rebinning it with a 
resolution of typically 3000 which does not lead to significant smearing for
NGC\,1068 linewidths. For those ranges affected by fringes, the single-valued 
spectra were defringed using the iterative sine fitting option of the aarfringe
module within the SWS Interactive Analysis.

The large number of observations required special treatment of redundant data.
In addition, observations from revolution 285 where apparently affected by a
slight ($\approx\/2-3\arcsec$?) pointing problem, which occured
occasionally in the earlier phase of 
the ISO mission. Since SWS beam profiles in some AOT bands are peaked and 
slightly offset with respect to the nominal pointing (A. Salama, 1999, 
priv. comm.), modest pointing offsets can cause 
noticeable flux losses in some AOT bands and resulting band mismatches. Such
mismatches were evident in revolution 285 band 3 data. Since most of the 
NGC 1068 mid-infrared flux comes from a small region (\cite{cameron93};
\cite{braatz93}; \cite{bock98}),
we corrected for this problem and the small scatter between other observations
by the following scaling procedure: For the SWS01 full spectrum obtained in
revolution 285, the individual AOT bands were scaled
to obtain both good match at band limits, and good agreement with the
overall flux level as estimated from our other SWS data and ground-based
photometry (\cite{lebofsky78}; \cite{rieke75}). At wavelengths below 
10$\mu$m photometry from different epochs should be used with great caution 
because of the known
variability (\cite{glass97}), our fluxes are however in good agreement with
the photometry of Glass for the ISO epoch. All other data were then scaled
to this SWS01 spectrum by the ratio inferred from the continuum flux densities.
We believe the final flux scale to be accurate within the 20-30\% typical for
SWS data (\cite{schaeidt96}).

Accurate wavelength calibration of the SWS grating spectrometer is central
for part of our line profile analysis, since shifts between lines in 
NGC\,1068 tend to be of the order 300\,km/s or less (\cite{marconi96}).
Valentijn et al. (1996) deduce an accuracy of $\sim$30km/s from extensive 
calibrations during the SWS performance verification phase. Since then, 
a slow secular drift in SWS wavelength calibration has been calibrated to 
similar accuracy. We have tested the 
wavelength calibration of the NGC\,1068 data, using identical calibration files
to analyze spectra of the planetary nebulae NGC\,7027 and NGC\,6543 taken close
to revolution 633 where some of the most important NGC\,1068 lineprofiles were
taken. We confirm the excellent accuracy from these observations, the largest
error not exceeding the value given by Valentijn et al. (1996). This test and
the good internal consistency of velocities measured in NGC\,1068 for 
different lines 
from the same species spanning most of the SWS wavelength range 
(e.g. H$_2$, see Table~\ref{tab:lineflux}) leads us to adopt an upper limit of 
50 km/s for any systematic errors in our wavelength scale, taking into
account a margin for mispointing.  

Two emission features, which were
tentatively detected in the preliminary analysis of Lutz et al. (1997)
could not be confirmed with the larger observational database and the
improved calibration. A broad
emission feature near 19$\mu$m, which might be interpreted as silicate
emission, is not confirmed with the new spectral response calibration.
An emission line at 28$\mu$m was identified as an unusually strong H$_2$ S(0) 
line. This identification 
was later found suspect
because of the line's larger width compared to the other H$_2$ lines
observed in NGC 1068.  
The line was not confirmed in deeper follow up observations.
Detailed inspection of the original data indeed
shows that it is an artifact of a highly unlikely coincidence of detector 
`glitches' at the expected wavelength of the S(0) transition.

\section{Results}
\label{sect:res}

The 2.4-45$\mu$m full spectrum of NGC\,1068 is displayed in 
Figure~\ref{fig:fullspec}. Some solid state features 
are superposed on the strong AGN-heated mid-infrared continuum.
These include 3.4$\mu$m C-H absorption, 9.6$\mu$m silicate absorption, and 
7.7, 8.6, 11.3$\mu$m `PAH' emission
(see Sturm et al (2000) for a discussion in conjunction with other SWS
spectra of galaxies).
Bright fine structure emission lines, mainly originating in the narrow
line region, are already visible in the full spectrum.
Figures~\ref{fig:ionspec} and \ref{fig:h2spec} show individual emission
lines. The display range is chosen to be $\pm$2500\,km/s around systemic 
velocity for recombination and
fine structure lines, and $\pm$1000\,km/s for the much narrower molecular 
lines. Throughout this paper, we adopt a systemic velocity of 1148km/s
(\cite{brinks97}). Good rest wavelengths are available for the observed 
transitions
from the literature and from recent ISO determinations (\cite{feuchtgruber97}).
Many lines were observed repeatedly, Figures~\ref{fig:ionspec} and 
\ref{fig:h2spec}
show only the best quality data. Table~\ref{tab:lineflux} lists the measured
line fluxes and limits, presenting averages of independent measurements with
higher weight given to better data. Table~\ref{tab:lineflux} includes 
also upper limits for some transitions (not shown in 
Figure~\ref{fig:ionspec}) that were observed with good enough signal-to-noise 
ratio. We list such limits for 
transitions from elements like Na or Ar where other ionization stages are
detected.

Fluxes of relatively narrow lines,
for example from H$_2$ and [\ion{Si}{2}], were measured by direct integration
of the continuum subtracted line profiles. The large linewidth makes this
procedure error prone for faint lines from the NLR, where continuum definition
is the main source of measurement error. 
The relative constancy of NLR line profiles
over a wide range of lower ionization potentials (see below) lead us to adopt
a different procedure to measure fluxes for the NLR lines: We derived a simple
two-gaussian template from the brightest NLR lines
([\ion{O}{4}] 25.89$\mu$m (note nearby [\ion{Fe}{2}]),
[\ion{Ne}{5}] 24.32$\mu$m, [\ion{Ne}{6}] 7.652$\mu$m) and used fits of this 
template (Figure~\ref{fig:allprof}) plus
a linear continuum to measure the fluxes of fainter NLR lines. The two
gaussian components have FWHM 333 and 1246\,km/s, peak ratio narrow/wide 
1.34, and the wider component is blueshifted by 100km/s. In fitting, we varied
only continuum flux and slope, total line flux, and total velocity. 
The fluxes measured this way
agreed very well ($\lesssim$10\%) with those determined by direct integration 
not only for the lines used to derive the template, but also for other bright
NLR lines like [\ion{Mg}{8}]\,3.028$\mu$m. The fit thus preserves the 
fluxes for bright lines and is preferable for faint lines
where continuum subtraction is the dominant source of error. 

For the blended lines of [\ion{Mg}{7}] and H$_2$ (0-0) S(7) near 5.5$\mu$m,
we list the fluxes resulting from a tentative gaussian fit using
two components for [\ion{Mg}{7}] and one component for H$_2$.
The uncertainty of the S(7) flux is particularly large,
up to a factor 2. The H$_2$ excitation diagram
(Figure~\ref{fig:excit}) in fact suggests that it may be overestimated.
Similar caution has to be applied to
the tentative flux listed for the [\ion{Fe}{2}] 25.99$\mu$m line. A slight
shoulder appears in the long wavelength wing of the [\ion{O}{4}] 
25.89$\mu$m transition, but its flux is very uncertain and may at best be good
enough to serve for consistency checks with other [\ion{Fe}{2}] lines. 
Feuchtgruber et al. (1997) discuss evidence that the lines of [\ion{Ar}{3}]
and [\ion{Mg}{7}] at 9.0$\mu$m are blended at the resolution of SWS.
We list only a total flux in Table~\ref{tab:lineflux}.

We detect a weak unidentified feature at rest wavelength about 7.555$\mu$m. 
If real, its width would suggest a NLR origin.
An instrumental origin due to an imperfection of the relative spectral response
function (RSRF) cannot be excluded but is unlikely since the RSRF shows
very little structure at that wavelength. Also, instrumental `ghosts' to 
strong SWS 
lines are not known at such a level. 
A possible interpretation of this feature is that it is a blueshifted
($\sim$3800 km/s) component of the nearby strong [\ion{Ne}{6}] line containing
$\sim$1.5\% of the total line flux. Residual instrumental fringing prevents
us from looking for analogous components near other strong lines such
as [\ion{Ne}{5}] or [\ion{O}{4}].
At this point, the 
feature must be considered as possibly real but without an obvious identification
by a line that is potentially strong in AGN spectra. Also, no line is seen at
this wavelength in archival ISO spectra of the high excitation planetary nebula
NGC\,6302.

We postpone a detailed discussion of the line profiles to 
section~\ref{sect:prof}.
Figures~\ref{fig:ionspec} and \ref{fig:h2spec} already suggest, however, 
that we are dealing with 
three distinct components: Fine structure transitions from species with
lower ionization potential $\gtrsim$ 40 eV (i.e. from [\ion{Ne}{3}] upwards)
have similar wide profiles and apparently originate in the NLR. 
Fine structure lines from lower ionization stages, in particular 
[\ion{Ne}{2}] and [\ion{S}{3}], are narrower and are most likely
contaminated by star formation within their beams. This limits their use
in modelling of the AGN-excited NLR spectrum (\cite{alexander00}) 
to upper limits rather than measurements. Inspection of
Figure~\ref{fig:ionspec} suggests that the starburst contribution 
strongly dominates the low excitation lines like [\ion{S}{3}] 33.48$\mu$m 
and [\ion{Si}{2}] 34.81$\mu$m which
are measured with the largest aperture. Lines measured with intermediate
apertures like [\ion{Ne}{2}] 12.81$\mu$m and in particular [\ion{S}{3}] 
18.71$\mu$m still show strong wings and will have a considerable NLR
contribution. The smallest line widths are measured
for transitions of molecular hydrogen.

\subsection{Density of the narrow line region}
\label{subsect:density}

The fine structure lines detected by SWS can be used to determine the density 
and, in conjunction with optical forbidden lines, the electron temperature
of the line emitting gas in the narrow line and coronal line regions. 
Differences in infrared
and optical lineprofiles (\S \ref{sect:prof}) discourage a 
determination of
electron temperatures from the integrated fluxes, which would not account
for significant variations in extinction across the NLR. A reliable average 
density can be determined, however, from the mid-infrared lines alone which
are insensitive to electron temperature and extinction variations. 
The contribution of starburst excitation to the density-sensitive forbidden 
lines can be estimated using the large line
width variation between the NLR and the circumnuclear ring of star formation
regions.

The most suitable NLR density diagnostic is provided by the ratio of the 
[\ion{Ne}{5}] transitions at 14.32 and 24.32$\mu$m. These lines cannot be
diluted significantly by circumnuclear star formation since they are undetected
in starburst galaxies (\cite{genzel98}). They were
observed with the same SWS aperture size and with good signal-to-noise.
Adopting the same atomic data as Alexander et al. (1999, see also their Figure
3 for diagrams of several density sensitive ratios) and an electron
temperature of 10000\,K, the observed [\ion{Ne}{5}] ratio of 1.39 corresponds to
an electron density $\rm n_e\sim 2000$\,cm$^{-3}$ in the region of the NLR
where species with lower ionization potential near 100eV prevail.

A seemingly discrepant result is obtained from the [\ion{S}{3}] transitions
at 18.71 and 33.48$\mu$m -- the observed ratio of 0.73 is consistent with
the low density limit and corresponds to $\rm n_e\lesssim 500$\,cm$^{-3}$.
But, Figure~\ref{fig:ionspec} shows the line profiles of the two transitions
to be quite different: [\ion{S}{3}] 18.71$\mu$m shows strong wide wings
and is apparently NLR-dominated with small starburst contamination. In 
contrast, the larger aperture of [\ion{S}{3}] 33.48$\mu$m collects more
emission from the starburst ring showing up as a strong narrow component
of the profile. If {\em all} 18.71$\mu$m emission were from a NLR at 
2000\,cm$^{-3}$, the NLR contribution to the 33.48$\mu$m flux would be
about 1/3, consistent with the weaker wings of this line. 

A similar problem may affect, to a lesser degree, the density sensitive
ratio of the [\ion{Ne}{3}] transitions at 15.55 and 36.01$\mu$m.
The observed ratio of 8.9 is lower than but probably still consistent 
with the low 
density limit ($\sim$12) which applies up to the $\rm n_e\sim 2000$\,cm$^{-3}$ derived
from [\ion{Ne}{5}]. The modest signal-to-noise ratio of the 36.01$\mu$m line
makes it impossible to use the line profile to assess starburst contamination 
in the large aperture. A crude estimate can be obtained assuming that the ratio
of {\em starburst} [\ion{Ne}{3}] 36.01$\mu$m and [S\,III] 33.48$\mu$m seen
additionally in the large aperture is 0.03--0.04 
as in the prototypical starburst M\,82 (\cite{foerster98}). Then, 
$\gtrsim$10\%\ of the 36.01$\mu$m line would be starburst contamination,
bringing the ratio closer to its low density limit value.
The density in NLR regions dominated by lower excitation species like
[\ion{S}{3}] and [\ion{Ne}{3}] hence appears consistent with that
for the higher excitation region containing [\ion{Ne}{5}].

With respect to the coronal line region, the observed ratio 0.55 of the 
[\ion{Si}{9}] lines at 2.584 and 3.936$\mu$m is close to its low
density limit which implies $\rm n_e\lesssim 10^6$\,cm$^{-3}$. 
The same limit is found for the Circinus galaxy (\cite{moorwood96}) and
NGC 4151 (\cite{sturm99}). Such a density limit is consistent with all
popular scenarios for coronal line formation in AGN except for origin in a 
very dense transition region between NLR and BLR. 
\section{Line profiles and the structure of the narrow line region}
\label{sect:prof}

Integrated emission lines profiles are an indirect tool to constrain the 
dynamical structure and extinction properties
of the narrow line region. Different lines probe different parts
of the NLR and the velocity field is generally far from uniform. 
With the advent of linear optical detectors, considerable effort was devoted
to studies of both forbidden and permitted optical line profiles in Seyfert 
galaxies. 
Although there is still no full
consensus among different studies of the NLR forbidden lines, the
emerging picture is as follows.\\
(1) The forbidden lines in most cases
show blue asymmetries in the sense of a sharper falloff to the red than to the
blue. Line centroids are blueshifted with respect to the systemic
velocity, whereas line peaks in high resolution spectra are close to systemic
velocity. 
This has been most thoroughly studied in moderate excitation species
including [\ion{O}{3}] 5007\AA\/
(e.g. \cite{heckman81}; \cite{vrtilek85}; \cite{whittle85a}; \cite{dahari88}) 
but holds also for the higher excitation
coronal lines (\cite{penston84}).\\
(2) Line widths and blueshifts often vary between different species 
observed in the same source. Line profiles appear
to be correlated with the ionization potential and/or the critical
density. There are indications, but no complete consensus, that the 
correlation with the critical density may be the fundamental one
(e.g. \cite{pelat81}; \cite{penston84}; \cite{whittle85b};
\cite{derobertis86}; \cite{appenzeller88}).

Various scenarios have been put forward to explain these
trends. Most of them invoke extinction to explain blue
asymmetries, and the most popular ones assume outflow in a dusty
NLR, with higher excitation species probably originating closer
to the central source in regions of higher velocity and obscuration.
It is obvious that observations of {\em infrared} NLR emission are a
powerful independent method to test such scenarios: near- and mid- infrared
lines suffer more than an order of magnitude less extinction than in the 
optical. The combination of optical and infrared data should hence elucidate
the role of dust obscuration.
Comparison of recombination line profiles in the optical with infrared ones
would be advantageous because of the relative insensitivity of recombination
line emissivities to local gas conditions. The line-to-continuum ratio of
recombination lines in the infrared is low, however, and better profiles
are obtained for coronal and fine structure lines which additionally 
cover a wide range of excitations.
Sturm et al. (1999) have presented a first such analysis using ISO-SWS
observations of NGC\,4151. On the basis of the similarity of optical and
infrared profiles, they ruled out the most simple scenario of an outflowing
NLR with pervasive dust, and suggested either a geometrically thin but
optically highly thick obscuring disk, or an intrinsic asymmetry of the NLR.

Because of the large flux and width of its `narrow' lines, NGC\,1068
is best suited for a line profile analysis at the modest resolving
power ($\sim$2000) of ISO-SWS. Optical line profiles have been observed at 
very high
resolving power by various groups (e.g. \cite{pelat80}; \cite{alloin83};
\cite{meaburn86}; \cite{veilleux91}; \cite{dietrich98})
and show complex, multi-peaked structure related to individual cloud 
complexes within the narrow line region of NGC\,1068. 
Marconi et al. (1996) have 
extended this work into the near infrared. At lower resolving power,
they do not discriminate the fine details of the best optical profiles
but derive the line centroids by Gaussian fits for a large set of near-infrared
and optical lines. They find that all optical and near-infrared high excitation 
lines are significantly blueshifted with respect to systemic velocity
($>$200km/s for lower ionization potential $\gtrsim$ 20eV).
They interpret this significant trend as a consequence of non-isotropic flows
or ionization patterns rather than selective extinction effects.

We extracted
line profiles for five high signal-to-noise SWS lines by subtracting
a linear continuum fitted outside 2500km/s from the line center and normalizing
to the peak of the line. These five lines originate in species spanning a wide 
range of excitation potentials ranging from 55 to 303eV and are shown in
Figure~\ref{fig:allprof}. The remaining uncertainty of these profiles is
dominated by noise for the high excitation lines of [\ion{Mg}{8}] and
[\ion{Si}{9}] and by continuum uncertainties for the other lines.
These could be both due to weak underlying real continuum features
(e.g. PAH near [\ion{Ne}{6}]) and due to residual fringing ([\ion{Ne}{5}]
and [\ion{O}{4}]). We do not show low excitation lines with significant 
starburst contribution, and the [\ion{Ne}{5}] 14.32$\mu$m and 
[\ion{Ne}{3}] 15.55$\mu$m  lines which are 
consistent with those shown in Figure~\ref{fig:allprof} but more uncertain
due to fringing. Brackett $\alpha$ has a much lower line to continuum ratio but
still good signal-to-noise ratio, and a line profile similar to 
the fine structure lines, as discussed by Lutz et al. (2000) in the context 
of putting limits on a broad line region contribution. 

For lines too faint to derive a good line profile, we fitted a single
gaussian plus linear continuum to derive at least a centroid velocity
(Table~\ref{tab:lineflux}). While
such a gaussian is not a good approximation to the intrinsic NLR profile,
we adopted it for simplicity and for consistency with the 
optical/near-infrared data of Marconi et al. (1996). We also fitted
the two-component NLR profile of Figure~\ref{fig:allprof} but do not list the
derived velocities since they agree with the simple gaussian fit except for
an offset that is constant within the uncertainties. From repeated observations
for some of these lines, we estimate an error of $\lesssim$50km/s. 
For lines with no velocity listed in 
Table~\ref{tab:lineflux}, we estimate that the uncertainty of deriving 
the centroid of a broad noisy line is too large to include it into an 
analysis of NGC\,1068. None of them, however, is discrepant by more than
$\approx$300km/s which would suggest misidentification. 

In the following subsections, we will derive a large aperture optical NLR
line profile for comparison with the ISO data, compare optical and infrared
profiles and centroid velocities, and interpret the differences found.

\subsection{A large aperture optical line profile}
\label{subsect:ksoprofile}

Mismatch between the typically
small optical apertures and the large mid-IR ones is important
when attempting to compare optical and mid-IR line profiles.
Datacubes from imaging spectroscopy would be ideal to extract optical line 
profiles matching the ISO apertures. At this point, however, published imaging
spectroscopy of NGC\,1068 is either limited in field size (\cite{pecontal97})
or lacks wavelength coverage: The datacube of Cecil et al. (1990) has
been obtained with 2600km/s total coverage in the [\ion{N}{2}] lines that are
additionally heavily blended with H$\alpha$, making it difficult to determine
the extent to which broad components are missing in their total line profile
(their Fig.~7).

We make use of two auxiliary large aperture optical spectra to address 
the problem of aperture mismatch: A 4000 to 7800\AA\/ spectrum from
Wise Observatory (WO, S. Kaspi 1999, priv. communication), providing good 
fluxes of the brightest lines 
in a $10\arcsec\times\/15\arcsec$ aperture (position angle 0\arcdeg), and a 
high spectral 
resolution Coud\'e Echelle spectrum from Karl Schwarzschild Observatory 
Tautenburg (KSO, E. Guenther 1999, priv. communication), providing a 
good [\ion{O}{3}] line profile (though not good fluxes) in a 
$6.8\arcsec\times\/15\arcsec$ aperture (mean position angle -28\arcdeg\/, 
varying during integration). In addition, we estimated relative 
emission line fluxes in our apertures by integrating over the
corresponding regions of a narrow band [\ion{O}{3}] map (R. Pogge, 
M.M. deRobertis, 1999, unpublished data).
Both the Wise spectrum and the [\ion{O}{3}] map confirm that ISO line fluxes 
of the NGC\,1068 NLR can be sensibly compared to smaller aperture 
optical data, since those already sample most of the flux in the narrow
line region. For example, a 4\arcsec\/ diameter aperture will already get 
$\approx$70\% of the flux in the ISO aperture.
The fluxes measured in the large Wise aperture for the brightest optical lines
agree within $\sim$30\% with published smaller aperture ones (e.g. 
\cite{shields75}; \cite{koski78}; \cite{marconi96}). 

Figure~\ref{fig:ksoveill} displays our KSO $6.8\arcsec\times\/15\arcsec$ 
aperture [\ion{O}{3}] 5007\AA\/ line profile in comparison to its 
$2.5\arcsec\times\/2.5\arcsec$ equivalent (\cite{veilleux91}). The line profile
changes induced by this more than tenfold increase in aperture area are 
relatively modest and fit expectations from high resolution longslit
spectroscopy. While the major components of Veilleux' spectrum are
well reproduced in the KSO data, their ratios differ somewhat leading to an 
overall slightly wider profile. This is fully consistent with observations
of relatively broad components over larger regions not sampled by Veilleux' 
aperture
(\cite{pelat80}; \cite{alloin83}; \cite{meaburn86}). The only feature
in the KSO profile not present in the Veilleux profile is an additional
narrow feature at or slightly redshifted from systemic velocity. This feature
almost certainly corresponds to the `velocity spike' in the NE region of the
NLR detected by many authors but seen perhaps most clearly in the data of
Meaburn \& Pedlar (1986). This feature is missed by Veilleux' aperture but
partly covered by the KSO data.

The KSO aperture is still about three times smaller in area than the SWS 
apertures through which the best fine structure line profiles have been taken.
The drop in [\ion{O}{3}] surface brightness with radius is so rapid
(e.g. Fig. 1 of Meaburn \& Pedlar 1986) that only modest differences 
in the total line profile are expected. An exception to this
is the NE region of the NLR  about 6\arcsec\/ from the nucleus
which was incompletely covered. The KSO slit orientation cannot
be chosen freely and was approximately aligned with the ISO
apertures but not with the NLR (PA -28\arcdeg\/
instead of PA $\approx$30\arcdeg\/), missing part of the NE end of the NLR. 
Long slit spectroscopy (e.g. \cite{meaburn86}) 
shows this NE region to be dominated by the narrow `velocity spike' near
systemic velocity which is already seen in the comparison of KSO and
Veilleux (1991) profiles. We hence expect this spike to be more
prominent in an optical line profile fully equivalent to the ISO aperture.
Integrating the [\ion{O}{3}] map over the ISO and KSO apertures we estimate 
a need to add $\sim$11\% to the KSO flux to account for the NE region and 
other low surface brightness emission near systemic velocity. We have taken
this into account by adding such a narrow (FWHM 150km/s) component 
to the KSO profile, and will use this 
in the following as basis of our optical line profile 
comparison (see also Figure~\ref{fig:ksoveill}).
Use of such a modified profile is supported by the spectrum of Pelat \&
Alloin (1980) which was obtained with a rotating longslit sweeping across the
NE region of the NLR, and showing a similar narrow spike (their component 5).

\subsection{Line profile variations}

The optical and infrared line profiles in NGC\,1068 differ strongly. This is most
evident in Figure~\ref{fig:opto4} which compares the profiles of 
[\ion{O}{4}] 25.89$\mu$m and [\ion{O}{3}] 5007\AA\/. We chose [\ion{O}{4}]
as the representative infrared line since it was observed with very good S/N 
and is close to 
[\ion{O}{3}] in lower ionization potential of the emitting species 
(55 vs. 35eV), ensuring origin in a similar region of the NLR. Critical 
densities ($1.0\times10^4$\,cm$^{-3}$ vs. $7.0\times10^5$\,cm$^{-3}$) match
less well than [\ion{Ne}{5}] or [\ion{Ne}{6}] would, but this is less relevant
given the low NLR density we have inferred (see \S \ref{subsect:density}). 

The optical [\ion{O}{3}] line profile is both blueshifted and broader than 
the infrared [\ion{O}{4}] line profile. 
There are however also significant similarities. Shoulders near
-900\,km/s and $\sim$350\,km/s are present in both profiles. 
Adopting different relative scalings (Figure~\ref{fig:opto4}), the 
impression arises that the two profiles in fact agree fairly well over
parts of their extent if the scaling is set properly. The main 
difference lies in {\em different relative strengths of blue wing, center,
and red wing}, the infrared profile having a stronger red wing and center.

The four highest quality infrared profiles are compared in 
Figure~\ref{fig:procomp} using a spread velocity scale. The most obvious
and significant variation is in [\ion{Mg}{8}] which is both broader and
more blueshifted than the lower excitation lines. The same is observed for
the more noisy [\ion{Si}{9}] line.
As already noted, the [\ion{O}{4}],[\ion{Ne}{5}], and [\ion{Ne}{6}] profiles 
with excitation energies ranging from 55 to 126eV are 
very similar but the overplot shows some variation in detail.
There is a minor shift in the narrow core of the [\ion{Ne}{6}] line which is,
however, not significant compared to the quoted systematic 
uncertainty. Comparing [\ion{Ne}{5}] to [\ion{O}{4}] taken
 from the same observation there is even less shift. Concerning
the broader wings, there are significant differences in addition to
the possible presence of [\ion{Fe}{2}] at $\approx$1100km/s in the 
[\ion{O}{4}] profile. There is a trend from [\ion{O}{4}] to [\ion{Ne}{6}] 
in the blue wing becoming stronger and the red wing fainter.

Such profile variations determine the line centroids derived from 
gaussian fits (Table~\ref{tab:lineflux}) to which we add a centroid
of 1015 km/s derived in the same way for our extrapolated large aperture 
optical [\ion{O}{3}]
profile. Anticipating that the shifts may reflect several
partially degenerate influences, we show in Figure~\ref{fig:centroids}
centroid velocities as a function of lower ionization potential, critical 
density, and extinction for the particular line. The extinction values are
relative and based on a preliminary ISO-based extinction curve for the center
of our Galaxy (\cite{lutz97a}). 

The trend observed in the ISO data for the velocity centroids as a function
of ionization potential
(Figure~\ref{fig:centroids}) is markedly different from the equivalent
dataset obtained in the optical and near-infrared (Figure~\ref{fig:centroids},
data taken from Marconi et al. 
1996). In both data sets the velocity centroids of the lowest excitation 
non-NLR lines
(H$_2$, [\ion{Fe}{2}]) are close to systemic. However, none of the ISO 
lines reach the large blueshifts observed consistently over a wide 
excitation range in 
the optical and near-infrared. 
The data sets are least discrepant at the high excitation end.
Here, [\ion{Si}{9}] 3.936$\mu$m is common to both sets and agrees
within the errors, though being somewhat less blueshifted in the ISO data. 
The discrepancy is largest at intermediate excitation
(20-200eV) where the optical/NIR lines are all strongly blueshifted while
the ISO lines only slowly deviate from systemic velocity as excitation
increases.

\subsection{Interpretation of the profile differences}

In addition to the cloud distribution and kinematics, line profiles 
reflect the emissivities of fine structure or forbidden lines 
in the narrow line region clouds, which are affected by many
parameters: ionization equilibrium, density, electron temperature, and 
extinction. 
If any of these
parameters vary among kinematically distinct structural components of the
NLR, line profile variations will result that in turn can help elucidate
the NLR structure. 
An important aspect is that some of these parameters
are partially degenerate in infrared datasets: Shorter wavelength (2-5$\mu$m)
lines which suffer higher extinction are also typically high excitation coronal 
lines with high critical densities, whereas the longer wavelengths are 
dominated by ions of lower excitation with transitions
with  lower critical densities.

Here we have assumed that the observed lines are emitted locally, with no 
contribution of scattered light. This is not strictly 
correct for part of the NGC 1068 narrow line region, in particular the NE
region (\cite{capetti95}; \cite{inglis95}). The impact 
of scattering on our analysis depends on the properties 
of the scatterer. If scattering is wavelength-independent, our profile
comparison is
unaffected since scattering would effectively only redistribute
the line emission spatially within our large apertures. If (dust) scattering
decreases with wavelength, shorter wavelength line profiles would be modified 
more strongly. We estimate the effect on the integrated line
profiles is not large, since the [\ion{O}{3}] polarisations measured by
Inglis et al. (1995) reach at most a few percent in regions that are in addition
minor contributors to the total flux, and since the considerable spatial 
variation in line profiles is not suggestive of scattered radiation
originating in a central source. 

 
It is unlikely that density
variations play a direct role in the profile variations.
Our density estimate of $\rm n_e\sim2000 cm^{-3}$
is too far below the relevant critical densities, making much higher
densities over a significant part of the NLR unlikely, which would be required 
to create the variations. Strictly speaking however, this estimate applies 
only to the moderately excited (100eV) gas, and our upper limit for the 
coronal region 
density is still consistent with densities higher than the critical densities 
of some lower excitation line. Strong collisional suppression of
part of some low critical density fine structure line profiles is also unlikely
from the similarity of their line centroids to that of the recombination line 
Brackett $\alpha$. We hence believe that the clear correlation of centroid
velocities with fine structure line critical density 
(Figure~\ref{fig:centroids}b) is mostly a secondary consequence of the 
correlation with ionization potential.

Variations in electron temperature can also affect the line profiles.
Emissivities of optical forbidden lines like [\ion{O}{3}] 5007\AA\/ strongly vary with
electron temperature while the infrared fine structure lines originate close
to the ground state and are much less sensitive to temperature. Hence, at least
those profile variations seen among the infrared lines (Figures
\ref{fig:procomp} and \ref{fig:centroids}) must be unrelated to temperature
fluctuations.
For the optical/IR profile variations there is a basic ambiguity
of an optical component being faint due to low electron temperature or
due to high extinction. Optical electron temperature determinations e.g.
from [\ion{O}{3}] 4363\AA\//5007\AA\/ are not available for the
various kinematic components of NGC\,1068. 
The line profiles of
the temperature insensitive recombination lines in the optical and IR
do not resemble each other, but rather follow the shapes of the forbidden 
optical and IR lines, respectively (\cite{veilleux91}, Fig.~\ref{fig:ionspec}). 
This is inconsistent with temperature variations being the main origin
of profile variations.

The exclusion of other factors suggests that ionization structure and
extinction are the main source of the observed variation of optical/IR line 
profiles.  Previous studies of the spatial and 
kinematical structure of the NGC\,1068 NLR (e.g. \cite{cecil90}, 
\cite{marconi96}) point to the existence of two spatial and
kinematical components. The first is
a strong ionization cone with associated blueshifted outflow, 
with the highest excitation species likely
concentrated towards the central and fastest part of this cone. The second
is an extended system of photoionized clouds closer to 
systemic velocity.
The correlation between the ISO centroid 
velocities and the ionization potential (Figure~\ref{fig:centroids}a)
fits this picture well. If the
relative importance of the fast outflow gradually increases towards high
excitation lines, the gradual centroid shift is easily explained.
However, the marked differences between optical [\ion{O}{3}] and 
infrared [\ion{O}{4}] profile (Figure~\ref{fig:opto4})
and the different optical and infrared centroid velocities at
similar ionization potential (Figure~\ref{fig:centroids}a, including data from 
Marconi et al. 1996) show that this picture must be incomplete.
We suggest that these remaining differences are due to extinction
variations across the NLR, with a general trend of higher exctinction in the
redshifted (SW) than in the blueshifted (NE) part. 
The optical profile can be
explained by an intrinsic profile similar to that of the IR lines
whose line center and red wing are reddened by a few magnitudes,
leading to the obscuration of about half of the total line flux.
A similar extinction pattern is suggested by the increase of polarization 
from the blue to the red wing of the {\em narrow} lines in the central
arcseconds (\cite{antonucci85}; \cite{bailey88}; \cite{inglis95}), 
attributed to absorption by aligned grains.
 
Considering that the optical profiles are modified by extinction, it is 
important to recognize that the less obscured near- and mid-infrared lines
remain blueshifted with respect to systemic velocity, high excitation ones
more strongly than lower excitation ones. Also, coronal line emission
observed in the little obscured near-infrared is still much stronger in 
the northeast cone than in the
southwest (\cite{thompson99}). This might be due to heavy 
extinction of an intrinsically symmetric NLR, or due to a real asymmetry.
No distinction can be made from the present data.
A trend with IR extinction is not clear in Figure~\ref{fig:centroids}c and
would be difficult to separate from the trends with ionization potential and 
critical density because of the mentioned degeneracy.
An intrinsically asymmetric NLR with outflow preferentially towards the 
observer is fully consistent with the observations of NGC\,1068. The difference
in strength of the two radio lobes (\cite{wilson83}) may be in support of
such an asymmetric scenario, although the relation between radio lobe flux and
NLR emission is certainly not simple.
Consistency with an asymmetric NLR was also noted for the
ISO spectra of NGC\,4151 (\cite{sturm99}). However, large randomly oriented AGN 
samples should not
show the preferential blueshift which is noted at least in optical 
samples. An alternative scenario of a geometrically small but 
optically highly thick screen (disk or torus) obscuring part of the receding 
NLR region was proposed by Sturm et al. (1999) for NGC 4151. Such a scenario
can also fit the NGC\,1068 data provided the screen obscures a 
relatively larger fraction of the coronal line region than of the
larger NLR which is dominated by medium excitation species. 
This is not implausible given the similar (arcsecond) spatial scale of
the central concentration of high column density molecular gas 
(e.g. \cite{tacconi94}) and the
coronal line region as mapped in [\ion{Si}{6}] (\cite{thompson99}; 
\cite{thatte00}). 
While studies of few individual sources will remain ambiguous, a search
for preferential shifts using high resolution near- or mid-infrared spectroscopy
should address this issue provided the sample of Seyferts is large enough and 
not biased by orientation of a putative asymmetric outflow, as may be the case
when identifying Seyferts in the optical.

Overall, the structure of the NGC\,1068 narrow line region seems to 
determine the optical/infrared line profiles via differences in weight of 
outflowing cone and extended components, and additional extinction 
variations that are significant for the optical wavelength range. 
The relative weight of the cone is larger for higher excitation species.
A tantalizing ambiguity remains that cannot
be resolved from a single source study: Is the NLR intrinsically one-sided
or is there a very high obscuration screen blocking also part of the IR
emission from our view?
\section{Molecular hydrogen emission}

NGC\,1068 was the first galaxy detected in the rovibrational transitions
of molecular hydrogen (\cite{thompson78}) and has been studied since in 
considerable detail both in these lines tracing fairly excited molecular
material (e.g. \cite{oliva90}; \cite{blietz94}; \cite{davies98}), and by 
millimeter wave interferometry tracing
colder components (e.g. \cite{tacconi94}; \cite{sternberg94}; 
\cite{helfer95}; \cite{schinnerer99}). The system of dense, warm 
cloud cores in the central few arcseconds inferred from the millimeter
studies calls for observations in the pure rotational transitions of molecular 
hydrogen, which trace gas of typically a few 100K, intermediate between the 
near-infrared and millimeter wave tracers.

Our SWS observations of these lines (Figure~\ref{fig:h2spec}, 
Table~\ref{tab:lineflux}) are summarized in the H$_2$ excitation diagram of
Figure~\ref{fig:excit}. The diagram has a curved shape suggestive of a 
mixture of temperatures, as expected from other
galaxies observed with ISO in the rotational lines of molecular hydrogen
(\cite{rigopoulou96}; \cite{sturm96}; \cite{valentijn96a}; \cite{kunze96};
\cite{spoon00}). 
The location
of the S(3) point slightly below the general trend suggests a moderate 
extinction towards this line whose wavelength is near the center of the silicate 
absorption feature ($\rm A_{9.6\mu\/m}\lesssim 1$). As noted earlier, the flux for 
the heavily blended S(7) line is very uncertain, so that the shorter wavelength
rovibrational lines of similar excitation will probably be a more 
trustworthy representation of the
excitation diagram at these upper level energies. 

While the higher rotational lines like S(5) and S(7) probe the same
excited but low mass component as the near-infrared rovibrational lines,
the bulk of the warm gas observed with ISO will reside in the component
traced by the S(1) line. In estimating its mass, we will assume a hydrogen
ortho/para (O/P) ratio in equilibrium at the local temperature
(\cite{sternberg99}). Combining the S(1) flux with the S(0) limit on
one hand and the S(3) detection on the other, the temperature of the S(1) 
emitting gas is found to lie in the range 140K$\le$T$\lesssim$375K, 
assuming the extinction
correction for S(3) is modest and aperture effects are minor. Because of the 
very steep temperature sensitivity
of the rotational line emissivities, the corresponding mass varies drastically,
between about $\rm 4\times\/10^6M_{\sun}$ for the 375 K case and 
$\rm 1.5\times\/10^8M_{\sun}$ for the 140 K case. For comparison with
other mass estimates, we will adopt $\sim$200\,K and
$\rm \sim\/2.5\times\/10^7M_{\sun}$ as a possible approximation
to the curvature of the excitation diagram. This mass would be of the
order 5\% of the total gas mass estimated from the 
CO interferometric map (\cite{helfer95} and priv. comm.) within the 
17$\mu$m ISO beam. Compared to a similar
estimate for the starburst galaxy NGC\,3256 (\cite{rigopoulou96}, 3\% for 150K
warm gas temperature), the fraction of warm gas and/or its temperature must
be higher, but not exceeding on average that inferred by Kunze et al. (1996) 
for the highly active star forming region in the `overlap region' of the 
antenna galaxies (8\% for 200K warm gas temperature). 

The NGC\,1068 molecular
hydrogen emission sampled by SWS may represent a mixture of relatively cool 
gas from the 15\arcsec\/ radius molecular ring partly covered by the longer 
wavelength apertures, and a warmer component from the unusually dense
and warm central few arcseconds (\cite{tacconi94}; \cite{blietz94}).
Knowing from the S(3) measurement that the extinction to the H$_2$ emitting 
region
cannot be very large, we can compare our 1-0 Q(3) flux to 1-0 S(1) fluxes
measured in smaller apertures (1.4 to 2.2$\times10^{-20}$W\,cm$^{-2}$;
\cite{blietz94}; \cite{thompson78}). Our Q(3) line flux falls in the middle 
of that
range. Since the intrinsic Q(3)/S(1) flux ratio is 0.7, the implication
is that the central few arcseconds dominate
at least for the more highly excited hydrogen lines, though there may be 
some extended contribution.
Some of the molecular hydrogen emission in NGC\,1068 may originate in
X-ray irradiated gas. We include in Table~\ref{tab:lineflux} upper limits
for two transitions of H$_3^+$ which have been proposed as a signature of
X-ray heated molecular gas (\cite{draine90}). We note, however, that these 
limits are not at all stringent and are fully consistent with even the
`high' end of H$_3^+$ flux expectations for X-ray illumination. Estimates are 
uncertain, see e.g. the much lower predictions of Maloney et al. (1996).  
 
The question can be raised whether the NGC\,1068 H$_2$ spectra in fact contain
a direct signature of a parsec-scale molecular torus. Krolik and Lepp (1989)
have modelled molecular line emission of such an X-ray illuminated torus,
predicting emission in some molecular hydrogen lines like (0-0) S(5) that
may under favorable conditions be detectable at ISO-SWS sensitivities.
For NGC\,1068 it is evident that larger scale emission may swamp any
possible torus emission, as already cautioned by Krolik and Lepp (1989).
The ISO data smoothly complement the near-infrared emission originating
in larger scale ($\sim$100pc) clouds, following an excitation diagram
plausibly ascribed to the same clouds. If any torus emission were present
at lower level, it may be difficult to discriminate from the larger scale
emission since, depending on
black hole mass and spatial scale, its velocity width could be very similar to
the larger scale emission. The widths of our H$_2$ rotational lines are 
consistent with those for CO observed on 100pc and larger scales, with
no evidence for other kinematic components. However, even if the rotational
emission does not trace a compact torus, it may still be excited
to a significant fraction by UV radiation, X-rays or shocks related to the AGN.

\section{Summary}

ISO-SWS spectroscopy provides the first detailed census of the mid-infrared
spectrum of the prototypical Seyfert 2 galaxy NGC\,1068. We
have detected 36 emission 
lines on top of the strong AGN-heated continuum. Most lines originate in the
NLR characterized by a density of $\sim$2000\,cm$^{-3}$.

We have compared the mid-infrared ISO line profiles with optical emission line
profiles produced in the NLR.
The line profiles are consistent with a model where the NLR is a 
combination of a highly ionized outflow and lower excitation extended emission,
with extinction significantly affecting the optical line profiles.
Remaining blueshift and asymmetry of the least obscured lines may reflect 
either intrinsic asymmetry of the NLR or an additional very high column density
obscuring component.

We detect strong emission from warm molecular hydrogen, which most likely 
originates on the
100pc to kpc scale, and which is also probed by emission in near-infrared and 
millimeter wave tracers
of molecular material. This emission masks any possible emission from a
putative parsec-scale molecular torus.

Companion papers use the SWS data to model the spectral energy distribution of 
the active nucleus, to put limits on emission from the obscured broad line 
region, and discuss the continuum and its features.  

\acknowledgments
We are grateful to Eike Guenther and Shai Kaspi for obtaining optical spectra
that were invaluable in the interpretation of the ISO spectra, and to 
Richard Pogge and  M.M. de Robertis for providing us with an
unpublished [\ion{O}{3}] image of NGC\,1068. 
SWS and the ISO Spectrometer
Data Center at MPE are supported by DLR (DARA) under grants 50 QI 8610 8 and
50 QI 9402 3.
We acknowledge support by the German-Israeli Foundation (grant 
I-0551-186.07/97).

%
%

\clearpage
\begin{table}
\caption{Journal of SWS observations of NGC\,1068}
\begin{tabular}{cccrc}
\tableline
\tableline
Revolution&Date       &AOT  &Duration&Slit Position angle\\
          &UT         &     &s       &\arcdeg\\
\tableline
285       &28-Aug-1996&SWS01&6538    &-11\\
285       &28-Aug-1996&SWS02&3212    &-11\\
605       &13-Jul-1997&SWS06&4080    &-23\\
633       &10-Aug-1997&SWS06&8410    &-16\\
643       &20-Aug-1997&SWS06&5710    &-13\\
788       &12-Jan-1998&SWS06&7766    &157\\
788       &12-Jan-1998&SWS02&1990    &157\\
792       &15-Jan-1998&SWS06&14642   &158\\
806       &29-Jan-1998&SWS06&11810   &162\\
818       &11-Feb-1998&SWS06&2134    &165\\
\tableline
\end{tabular}
\label{tab:listobs}
\end{table}

\clearpage
\renewcommand{\baselinestretch}{1.0}
\begin{table}
\caption{NGC\,1068 emission lines measured with ISO-SWS}
\scriptsize
\begin{tabular}{lrrrccr}
\tableline
\tableline
Line   &$\lambda_{\rm rest,vac}$&E$_{\rm Ion}$\tablenotemark{a}
                                     &Flux &Meth\tablenotemark{b}
                                          &Aperture&Velocity\tablenotemark{c}\\
                &$\mu$m        &eV   &10$^{-20}$\,W\,cm$^{-2}$&
                                             &$\arcsec\times\arcsec$&km/s\\
\tableline
H$_2$ (1-0) Q(3)& 2.424        &  0.0&  1.7&i&14$\times$20&1140\\
{}[Si\,IX]      & 2.584        &303.2&  3.0&f&14$\times$20&    \\
Br $\beta$, H$_2$ (1-0) O(2)&2.626&13.6&4.1\tablenotemark{d}&f&14$\times$20&\\
{}[Mg\,VIII]    & 3.028        &224.9& 11.0&f&14$\times$20&1000\\
{}[Ca\,IV]      & 3.207        & 50.9&  1.3&f&14$\times$20&    \\
H$_2$ (1-0) O(5)& 3.235        &  0.0&  0.8&i&14$\times$20&1120\\
Pf $\gamma$     & 3.741        &13.6&$<$4.0&u&14$\times$20&    \\
{}[Si\,IX]      & 3.936        &303.2&  5.4&f&14$\times$20& 950\\
Br $\alpha$     & 4.052        & 13.6&  6.9&f&14$\times$20&1110\\
H$_3^+1,2,3_{+1}\rightarrow3,3$&4.350&0.0&$<$3.0&u&14$\times$20&    \\
{}[Mg\,IV]      & 4.487        & 80.1&  7.4&f&14$\times$20&    \\
{}[Ar\,VI]      & 4.530        & 75.0& 15.0&f&14$\times$20&    \\
Pf $\beta$      & 4.654       &13.6&$<$10.0&u&14$\times$20&    \\
{}[Na\,VII]     & 4.685      &172.1&$<$11.0&u&14$\times$20&    \\
H$_2$ (0-0) S(9)& 4.695        & 0.0&$<$3.5&u&14$\times$20&    \\
{}[Fe\,II]      & 5.340        &  7.9&  5.0&i&14$\times$20&1120\\
{}[Mg\,VII]     & 5.503        &186.5& 13.0&s&14$\times$20&    \\
H$_2$ (0-0) S(7)& 5.511        &  0.0&3.5::&s&14$\times$20&    \\
{}[Mg\,V]       & 5.610        &109.2& 18.0&f&14$\times$20&    \\
H$_2$ (0-0) S(5)& 6.910        &  0.0&  5.9&i&14$\times$20&1180\\
{}[Ar\,II]      & 6.985        & 15.8& 13.0&i&14$\times$20&    \\
{}[Na\,III]     & 7.318        & 47.3&  5.8&f&14$\times$20&    \\
Pf $\alpha$     & 7.460        &13.6&$<$3.0&u&14$\times$20&    \\
Unidentified    &$\sim$7.555   &  ---&  1.8&f&14$\times$20&    \\
{}[Ne\,VI]      & 7.652        &126.2&110.0&f&14$\times$20&1030\\
{}[Fe\,VII]     & 7.815        & 99.1&  3.0&f&14$\times$20&    \\  
{}[Ar\,V]       & 7.902       &59.8&$<$12.0&u&14$\times$20&    \\
H$_2$ (0-0) S(4)& 8.025        &  0.0&  3.5&i&14$\times$20&1130\\
{}[Na\,VI]      & 8.611      &138.4&$<$16.0&u&14$\times$20&    \\
{}[Ar\,III]/[Mg\,VII]&8.991&27.6/186.5&25.0&f&14$\times$20&    \\
{}[Fe\,VII]     & 9.527        & 99.1&  4.0&f&14$\times$20&    \\
H$_2$ (0-0) S(3)& 9.665        &  0.0&  6.0&i&14$\times$20&1130\\
{}[S\,IV]       &10.511        & 34.8& 58.0&f&14$\times$20&    \\
H$_2$ (0-0) S(2)&12.279        & 0.0&$<$8.0&u&14$\times$27&    \\
{}[Ne\,II]      &12.814        & 21.6& 70.0&i&14$\times$27&1080\\
{}[Ar\,V]       &13.102       &59.8&$<$16.0&u&14$\times$27&    \\
{}[Ne\,V]       &14.322        & 97.1& 97.0&f&14$\times$27&1020\\
{}[Ne\,III]     &15.555        & 41.0&160.0&f&14$\times$27&1090\\
H$_3^+5,0\rightarrow4,3$&16.331&0.0&$ <$6.0&u&14$\times$27&    \\
H$_2$ (0-0) S(1)&17.035        &  0.0&  7.7&i&14$\times$27&1120\\
{}[Fe\,II]      &17.936        &7.9&$<$10.0&u&14$\times$27&    \\
{}[S\,III]      &18.713        & 23.3& 40.0&i&14$\times$27&1120\\
{}[Ne\,V]       &24.318        & 97.1& 70.0&f&14$\times$27&1060\\
{}[O\,IV]       &25.890        & 54.9&190.0&f&14$\times$27&1100\\
{}[Fe\,II]      &25.988        &  7.9&8.0::&s&14$\times$27&    \\
H$_2$ (0-0) S(0)&28.219        & 0.0&$<$2.5&u&20$\times$27&    \\
{}[S\,III]      &33.418        & 23.3& 55.0&i&20$\times$33&1110\\
{}[Si\,II]      &34.815        &  8.2& 91.0&i&20$\times$33&1110\\
{}[Ne\,III]     &36.014        & 41.0& 18.0&f&20$\times$33&    \\
\tableline
\end{tabular}
\label{tab:lineflux}
\tablenotetext{a}{Lower ionization potential of the stage leading to the
transition}
\tablenotetext{b}{Method of flux measurement: f = fit of double-gaussian NLR 
 profile, 
 i = direct integration, u = 3$\sigma$ upper limit assuming a line width of
 1000km/s (ions) or 300km/s (molecules), s = special fit. See also text.}
\tablenotetext{c}{Heliocentric velocity determined from fitting a single 
gaussian plus continuum to the -- sometimes complex -- line profile.}
\tablenotetext{d}{Brackett $\beta$ is blended with H$_2$ (1-0) O(2). From the
(1-0) Q(3) and (1-0) O(5) fluxes, we estimate a (1-0) O(2) contribution of 
$\sim0.4\times\/10^{-20}$\,W\,cm$^{-2}$ to the total flux}
\end{table}

\normalsize
\renewcommand{\baselinestretch}{1.6}

%
%

\clearpage

\figcaption[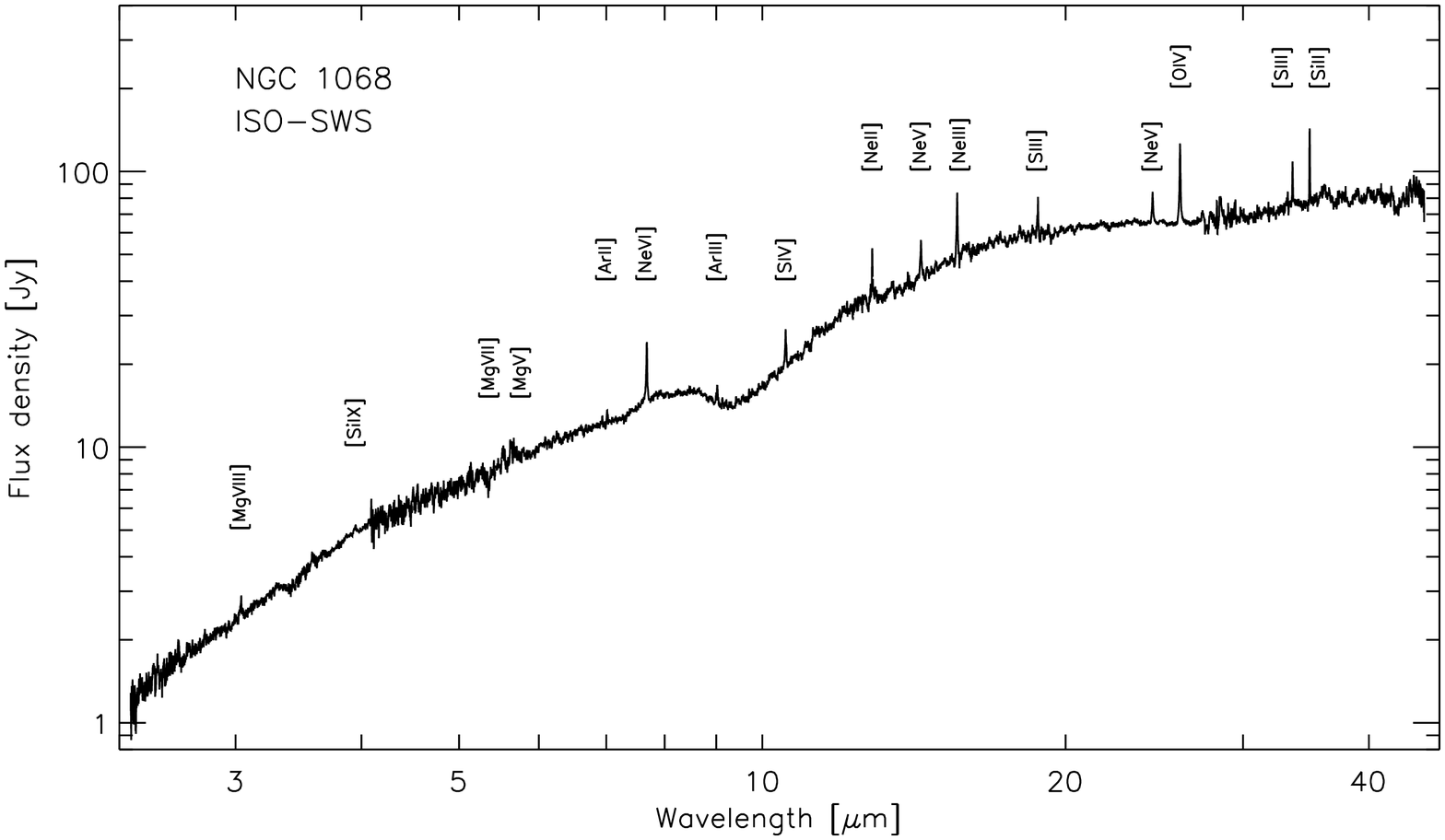]{Complete 2.4--45$\mu$m ISO-SWS spectrum of 
NGC\,1068. Some of the brightest emission lines are indicated.
\label{fig:fullspec}}

\figcaption[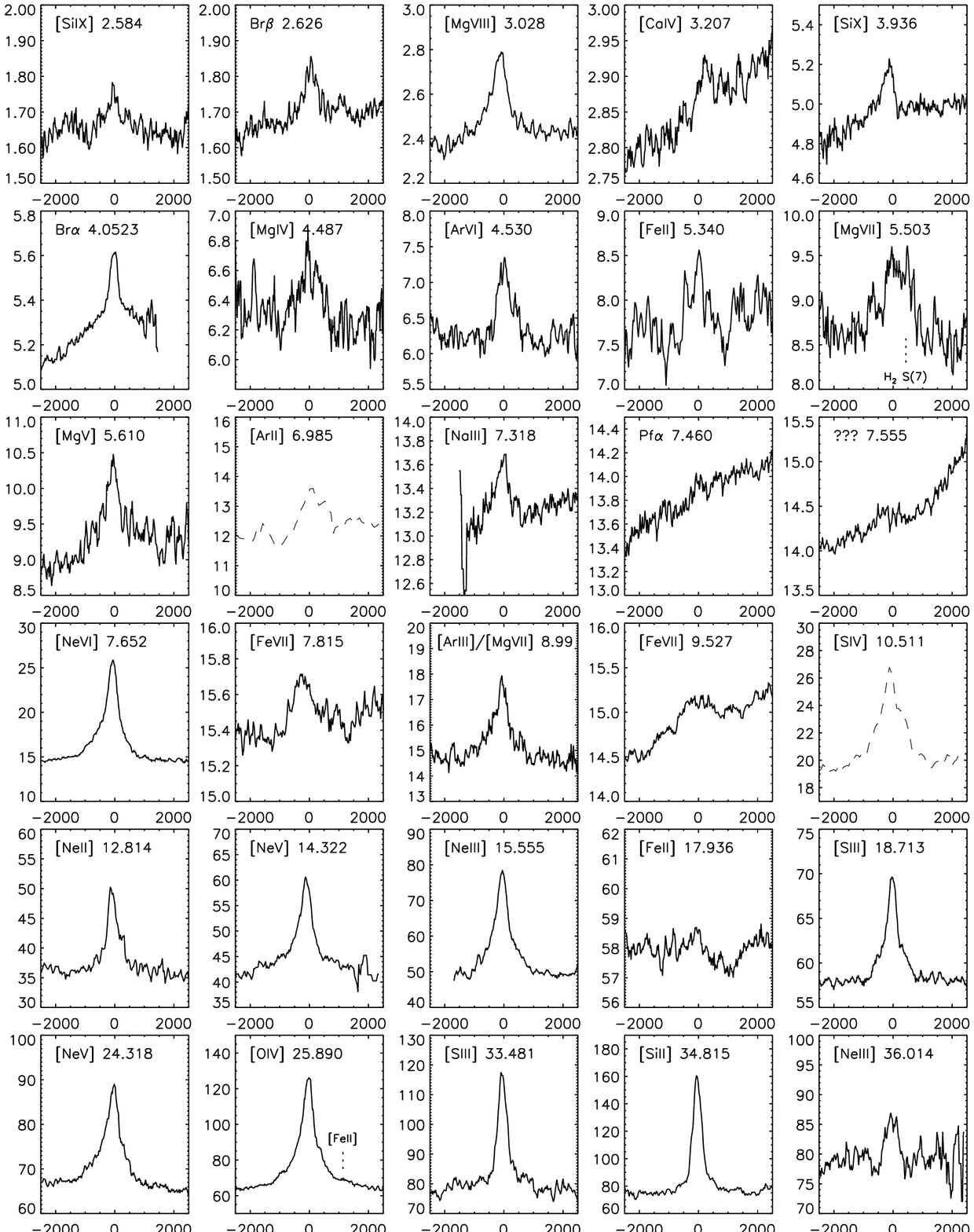]{ISO-SWS spectra of lines emerging in the ionized
medium of NGC\,1068. Flux densities in Jy are shown for a range of 
$\pm$2500\,km/s around systemic velocity. Most lines originate in the narrow 
line region but some low excitation 
lines have a significant starburst contribution. The two lines shown dashed
were observed in SWS01 mode, all others in full resolution SWS06 mode.
\label{fig:ionspec}}

\figcaption[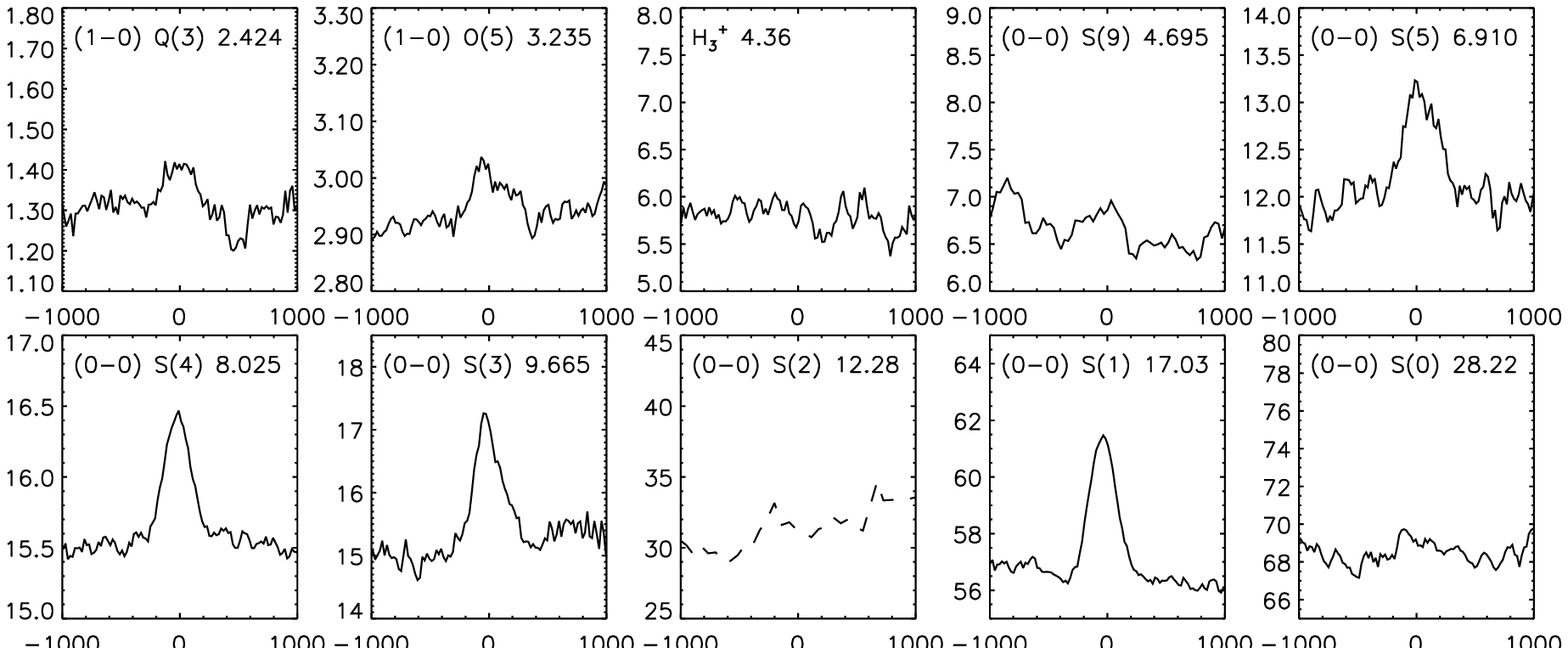]{ISO-SWS spectra of molecular transitions in 
NGC\,1068. Flux densities in Jy are shown for a range of 
$\pm$1000\,km/s around systemic velocity. The line shown dashed
was observed in SWS01 mode, all others in full resolution SWS02 or SWS06 modes.
\label{fig:h2spec}}

\figcaption[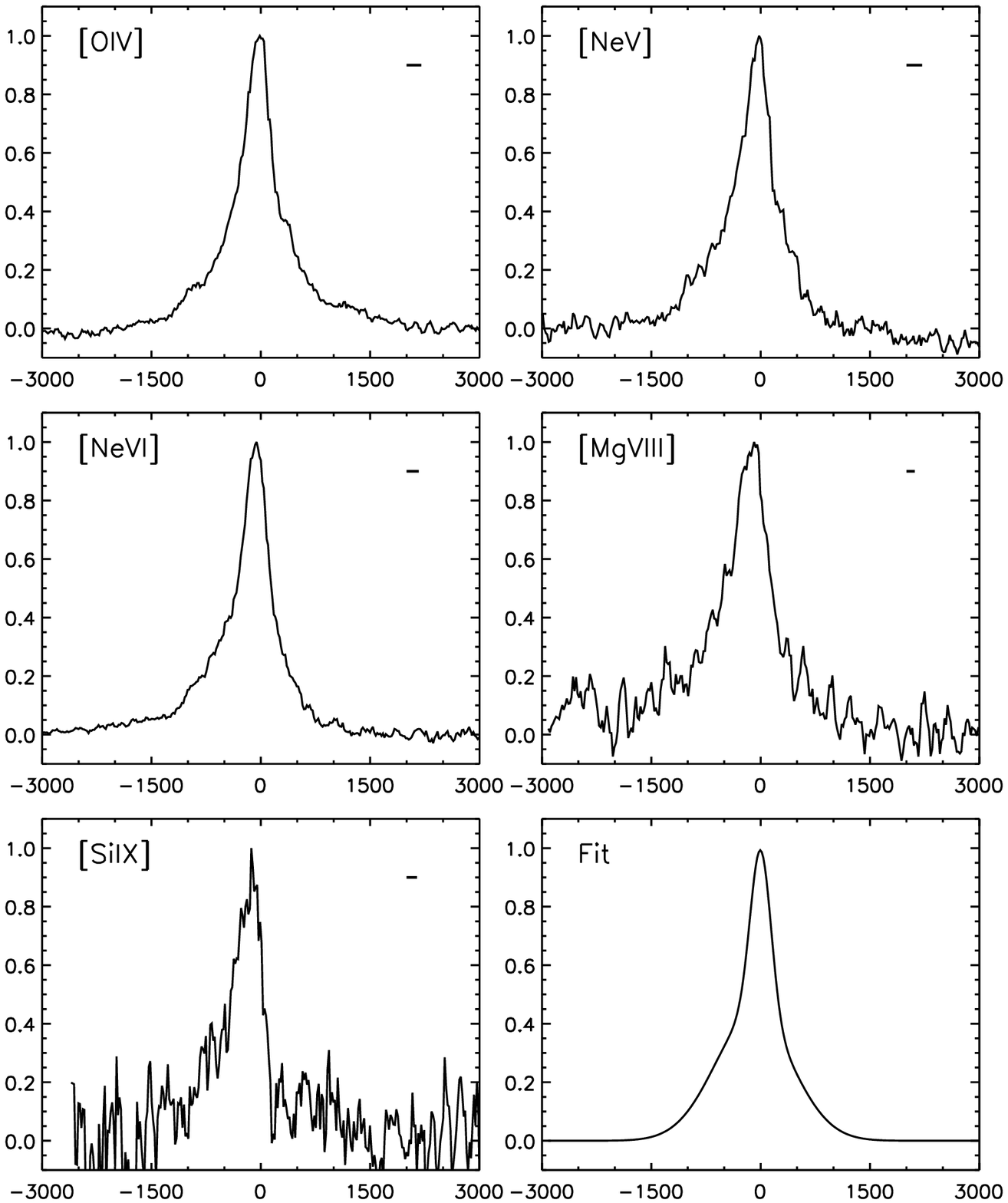]{Normalized line profiles for five high signal
to noise fine structure lines in NGC 1068, covering a range of lower 
ionization potentials from 55 to 303eV. The velocity scale is with respect 
to the heliocentric systemic velocity of 1148\,km/s. Short lines in the upper
right part of the panels indicate the SWS resolution at that wavelength.
In the lower right panel, we show a combination of two gaussians which
provides a reasonable approximation to the mid-infrared NLR line profiles 
over this wide range of ionization potentials. We have used fits of such 
a profile to measure {\em fluxes} of fainter lines.
\label{fig:allprof}}

\figcaption[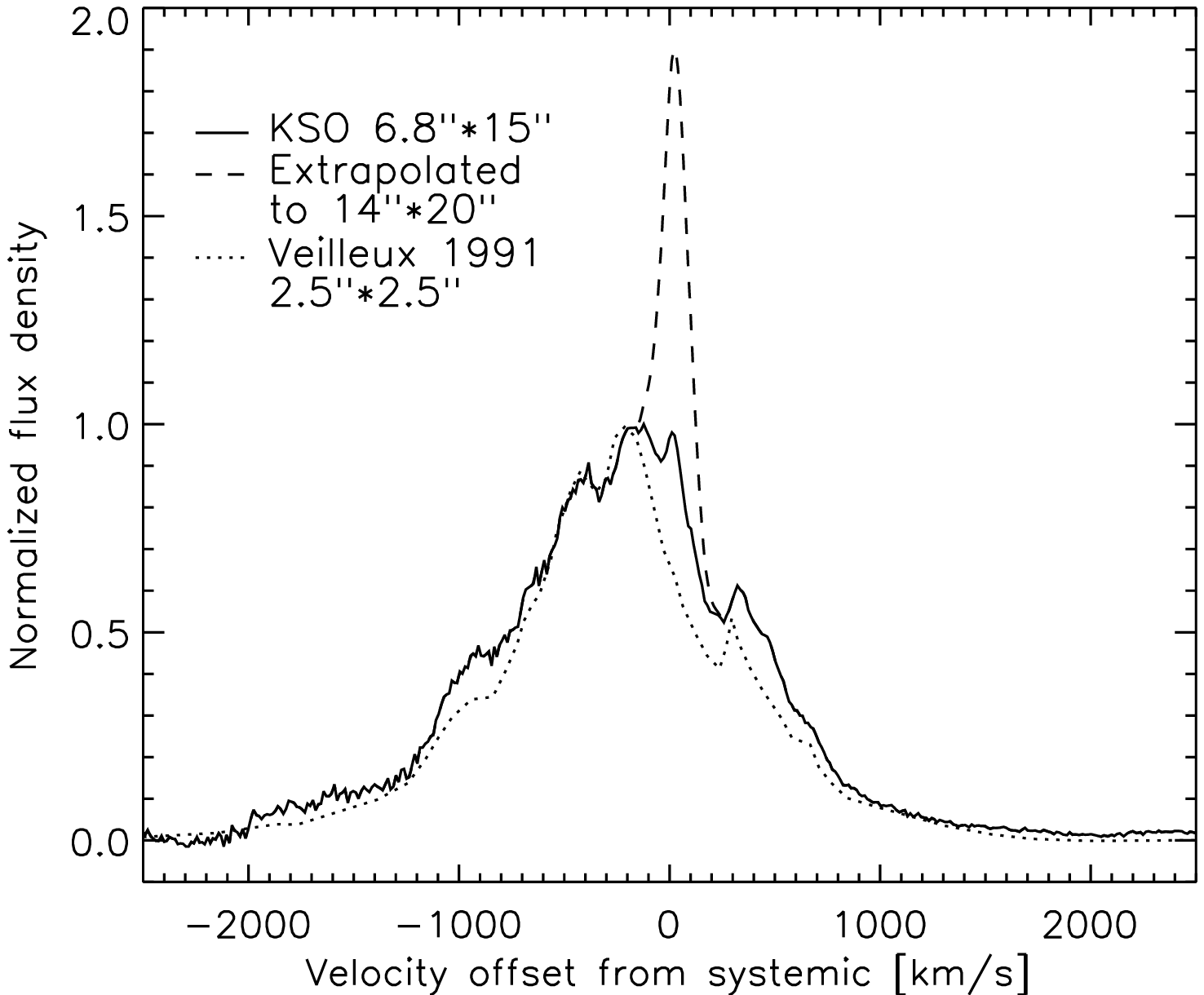]{Comparison of the large aperture [\ion{O}{3}] 
5007\AA\/ profile obtained at Karl Schwarzschild Observatory with the smaller
aperture one of Veilleux (1991). The velocity scale is with respect to the 
heliocentric systemic velocity of 1148\,km/s - note that this
differs from the original figure of Veilleux (1991). The dashed line indicates
a suggested extrapolation of the KSO spectrum to the larger ISO aperture
of 14\arcsec$\times$20\arcsec, taking into account extended narrow 
emission near systemic velocity, mainly from the NE region of the NLR.
See text for details.
\label{fig:ksoveill}}

\figcaption[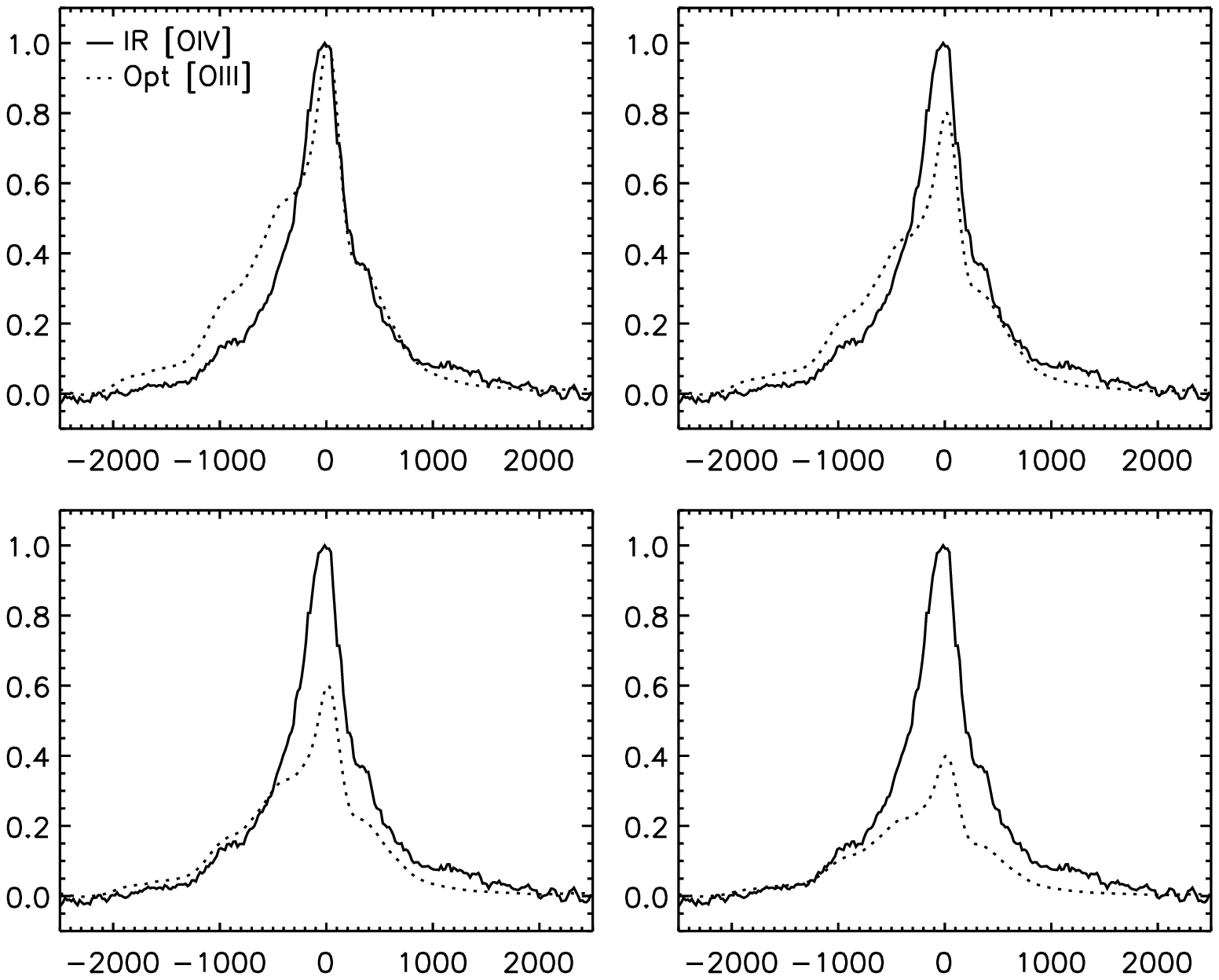]{Comparison of the infrared [O\,IV] and
optical [O\,III] line profiles. The peak of the optical profile
is normalized to 1, 0.8, 0.6, and 0.4 times the peak of the infrared
profile in the four panels.
The optical profile is the modified KSO profile of Figure 5
smoothed to the SWS resolution at the wavelength of [\ion{O}{4}]. While many 
structures are present in both optical and infrared profile, there 
are pronounced differences in the relative strengths of center and blue/red
wings. 
\label{fig:opto4}}

\figcaption[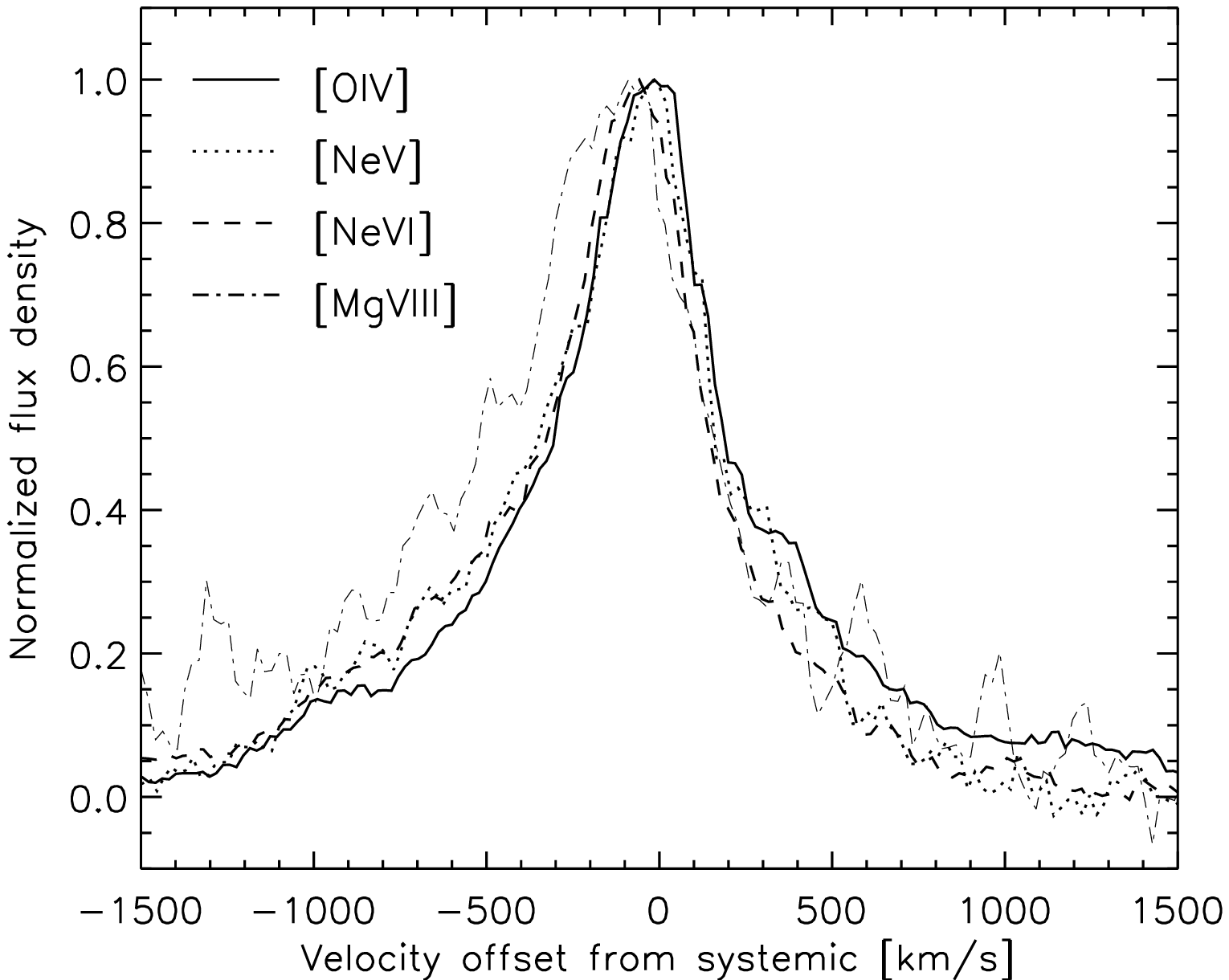]{Direct comparison of the four highest 
signal-to-noise fine structure line profiles. The velocity scale is with 
respect to the heliocentric systemic velocity of 1148\,km/s.
\label{fig:procomp}}

\figcaption[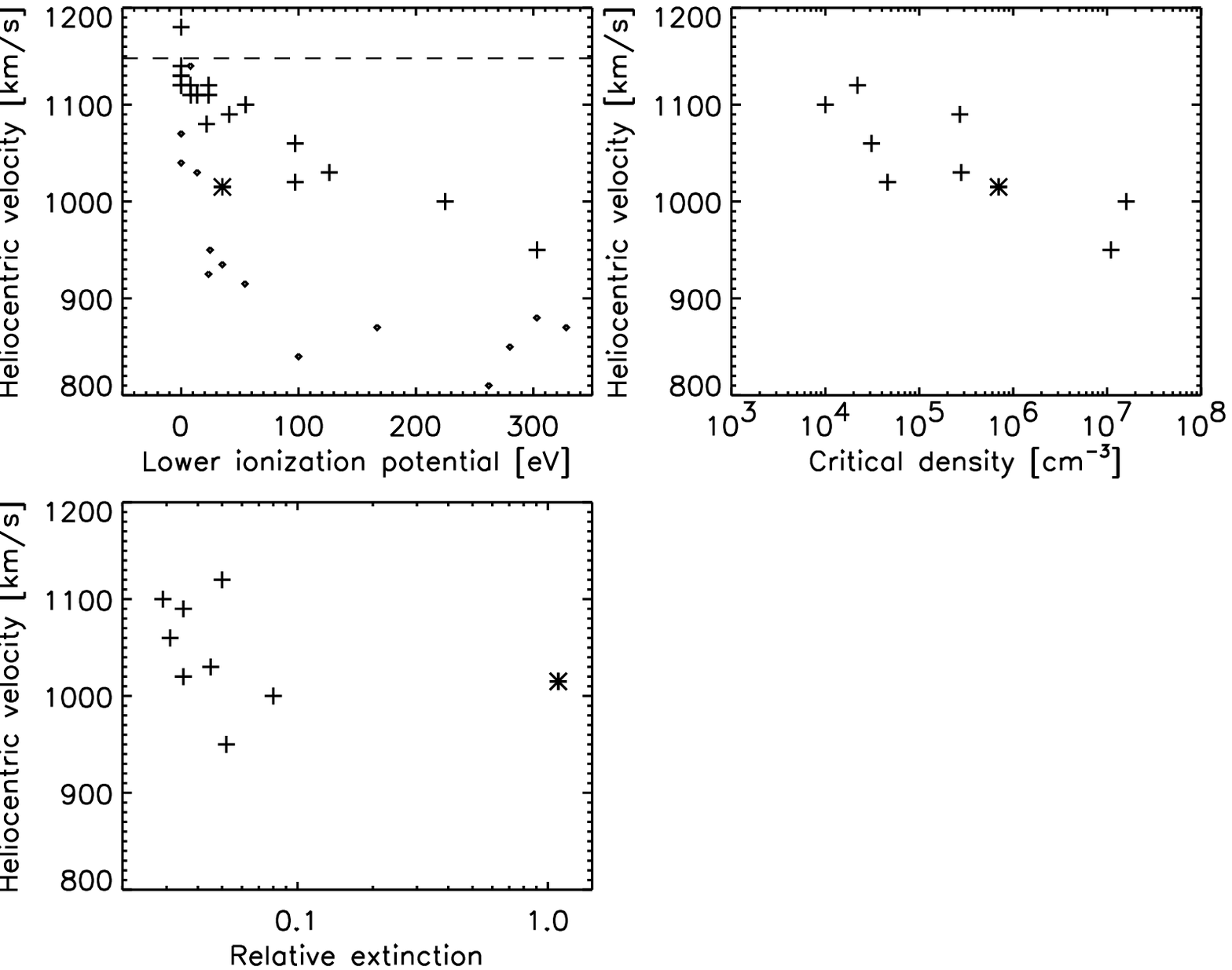]{Centroid velocities for NGC\,1068 emission 
lines, derived from fits of a single gaussian. Systemic velocity is indicated
by the dashed line in the upper left panel. The correlation with ionization
potential includes all ISO lines with a velocity listed in 
Table~\ref{tab:lineflux} (crosses) with the addition of the optical [\ion{O}{3}] line
(shown as an asterisk in all graphs). For comparison with the ISO
mid-infrared results, centroid velocities of optical/near-infrared
lines as derived by Marconi et al. (1996) are shown as small diamonds in the
upper left panel. The graphs showing velocity as a function
of critical density and of extinction are restricted to lines dominated by the
NLR according to the line profile. The velocities shown are accurate to
$\lesssim$50km/s. 
\label{fig:centroids}}

\figcaption[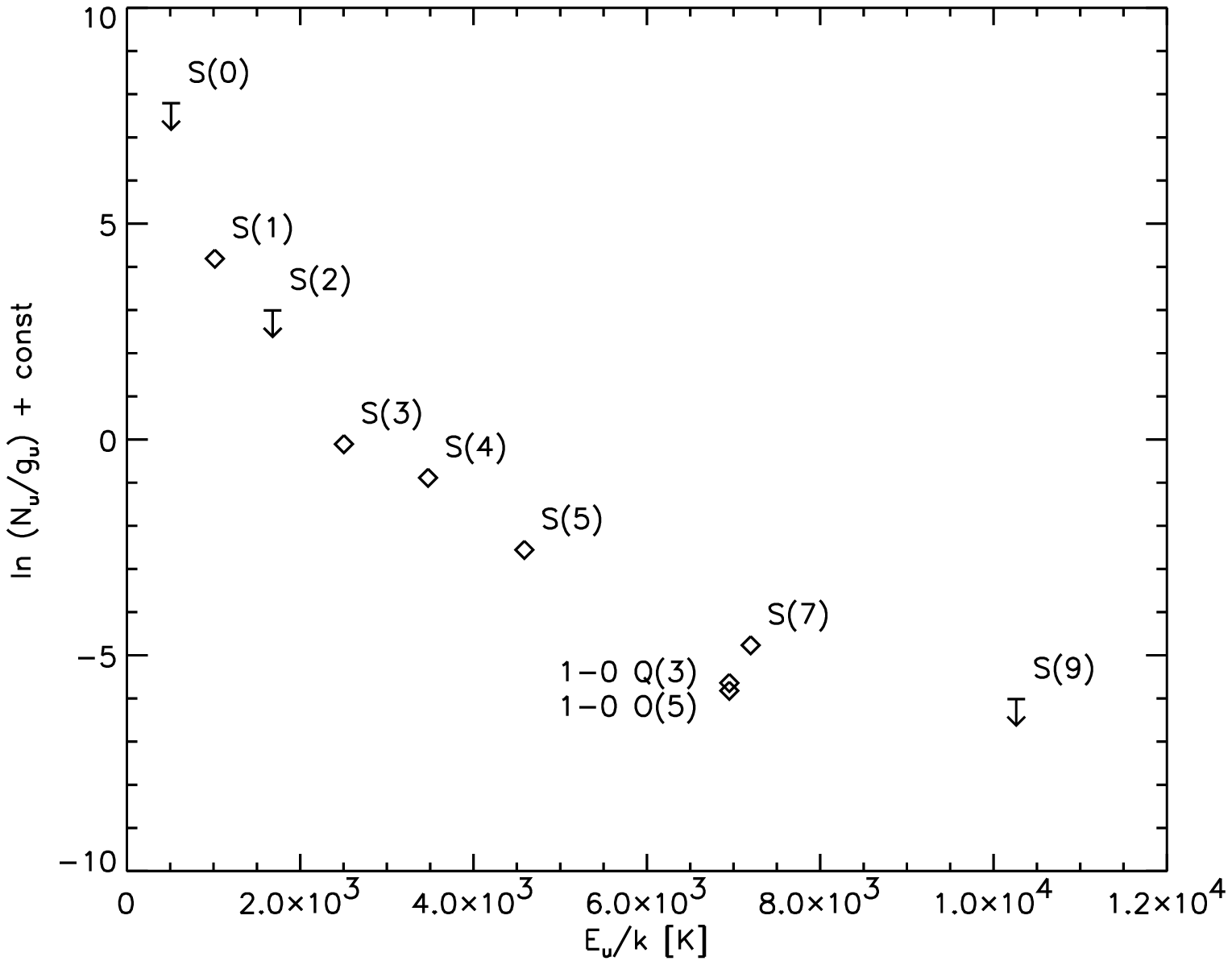]{
Excitation diagram for the molecular hydrogen lines measured with SWS in 
NGC\,1068.
\label{fig:excit}}

%
%
\clearpage

\plotone{fig_fullspec.eps}

\clearpage

\plotone{fig_ionspec.eps}

\clearpage

\plotone{fig_h2spec.eps}

\clearpage

\plotone{fig_allprof.eps}

\clearpage

\plotone{fig_ksoveill.eps}

\clearpage

\plotone{fig_opto4.eps}

\clearpage

\plotone{fig_procomp.eps}

\clearpage

\plotone{fig_centroids.eps}

\clearpage

\plotone{fig_excit.eps}

\end{document}